\documentclass[journal]{IEEEtran}
\usepackage{xcolor,soul,framed} 

\colorlet{shadecolor}{yellow}
\usepackage[pdftex]{graphicx}
\graphicspath{{../pdf/}{../jpeg/}}
\DeclareGraphicsExtensions{.pdf,.jpeg,.png}

\usepackage[cmex10]{amsmath}
\usepackage{array}
\usepackage{mdwmath}
\usepackage{mdwtab}
\usepackage{eqparbox}
\usepackage{url}
\usepackage{svg}

\begin{document}

\bstctlcite{IEEEexample:BSTcontrol}
    \title{Picosecond synchronization system for quantum networks}
     

\author{Raju Valivarthi, 
Lautaro Narv\'{a}ez,
Samantha I. Davis, 
Nikolai Lauk, 
Cristi\'{a}n Pe\~{n}a, 
Si Xie, 
Jason P. Allmaras,
Andrew D. Beyer, 
Boris Korzh, 
Andrew Mueller,
Mandy Rominsky,
Matthew Shaw, 
Emma E. Wollman, 
Panagiotis Spentzouris, 
Daniel Oblak, 
Neil Sinclair and 
Maria Spiropulu
\thanks{R. Valivarthi, L. Narv\'{a}ez, S.I. Davis, N. Lauk and M. Spiropulu are with the Division of Physics, Mathematics and Astronomy, and the Alliance for Quantum Technologies, California Institute of Technology, USA}
\thanks{S. Xie and C. Pe\~{n}a are with the  Division of Physics, Mathematics and Astronomy,  the Alliance for Quantum Technologies, California Institute of Technology and the Fermi National Accelerator Laboratory, USA}
\thanks{A. Mueller is with the Division of Engineering and Applied Science,  the Alliance for Quantum Technologies and the Jet Propulsion Laboratory, California Institute of Technology, USA}
\thanks {J.P. Allmaras, A.D. Beyer, 
B. Korzh, M. Shaw  and E. E. Wollman are with the Jet Propulsion Laboratory, California Institute of Technology, USA}
\thanks{M. Rominsky and P. Spentzouris are with the Fermi National Accelerator Laboratory, USA }
\thanks{D. Oblak is with the Institute for Quantum Science and Technology, and Department of Physics \& Astronomy, University of Calgary, Canada}
\thanks{N. Sinclair is with the John A. Paulson School of Engineering and Applied Sciences, Harvard University and the Alliance for Quantum Technologies, California Institute of Technology, USA}
}

\maketitle

\begin{abstract}
The operation of long-distance quantum networks requires photons to be synchronized and must account for length variations of quantum channels.
We demonstrate a 200 MHz clock-rate fiber optic-based quantum network using off-the-shelf components combined with custom-made electronics and telecommunication C-band photons.
The network is backed by a scalable and fully automated synchronization system with ps-scale timing resolution. 
Synchronization of the photons is achieved by distributing O-band-wavelength laser pulses between network nodes.
Specifically, we distribute photon pairs between three nodes, and measure a reduction of coincidence-to-accidental ratio from 77 to only 42 when the synchronization system is enabled, which permits high-fidelity qubit transmission.
Our demonstration sheds light on the role of noise in quantum communication and represents a key step in realizing deployed co-existing classical-quantum networks.
\end{abstract}


\IEEEpeerreviewmaketitle

\section{Introduction}

\begin{figure}[h!]
  \begin{center}
  \includegraphics[width=0.5\textwidth]{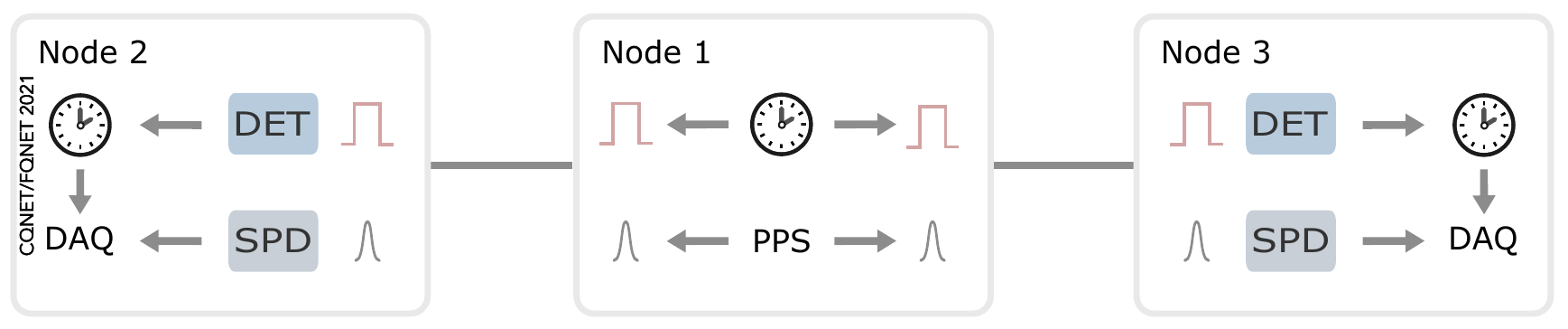}\\
  \caption{Concept of a clock distribution system for a three-node quantum network. An oscillator (clock face) is used to generate pulses (top hats) at a central node (node 1) that are distributed to end nodes (nodes 2 and 3) by fiber channels (grey lines) where they are detected (DET) and used to lock the phase of clocks at the end nodes. Simultaneously, light (Gaussians) from a photon pair source (PPS) at the central node is directed into the same fiber towards single photon detectors (SPDs) at the end nodes. Data acqusition (DAQ) systems record the arrival times of the photons with respect to the phase of the clocks at the end nodes, thereby ensuring the clocks are synchronized with the photons.
  }
  \label{fig:concept}
  \end{center}
\end{figure}

\IEEEPARstart{L}{ong-distance} quantum networks require distribution of qubits encoded into individual photons.
For deployed networks, photons must be transmitted using low-loss, high-bandwidth, and practical channels, such as fiber optics cables.
Ideally, photons must also be generated and detected at high rates, for instance using modulated lasers and high-timing resolution (low-timing jitter) nanowire detectors, respectively.
The realization of such networks is at odds with environment-induced variations in the length of fiber optics cables.
Since photons are identified by recording their times of generation and detection, each with respect to a local (node-based) clock, such variations can lead to misidentification of photons.
To avoid this, the variations can be accounted for by adjusting the phases of each local clock based on a centrally located primary clock. 
This is accomplished by distributing strong optical pulses to the nodes, as conceptualized by the diagram in Fig. \ref{fig:concept}.
These pulses can be distributed either separately in parallel fibers or jointly through the same fiber that is carrying the single-photon level quantum signal. This clock distribution method, which has been exploited in previous quantum networking demonstrations \cite{tanaka08, tang2015mdiqkd, valivarthi2016quantum,sun2016quantum, valivarthi2019measurement, williams2021qkdandsync}, allows synchronization between all local clocks, and hence identification of photons throughout the network.
Importantly, this method enables operation of the network at a high clock-rate and the possibility to perform linear-optic Bell-state measurements based on two-photon interference, which requires precise synchronization of photons.
Using a single fiber for both the synchronization and quantum signal makes better use of the limited optical fiber infrastructure that can be employed for quantum communication. The challenge with this setting is to ensure that the strong optical pulses used for synchronization do not introduce noise that reduces the fidelity of the transmitted qubits.
The leading source of noise in optical fiber channels is due to off-resonant Raman scattering of the clock pulses, which produces significant red-shifted light \cite{choi10ramanscatteringQKD}.
Typical methods to mitigate this involve strong temporal and spectral filtering \cite{patel2012coexistence} or using photons that are blue shifted from the clock pulses \cite{tanaka08}.
Accordingly, there is little work to investigate the role of the Raman noise if photons are red-shifted from the clock pulses and both are in the same fiber, in particular if the photons and the clock pulses are at wavelengths within the standard fiber telecommunication windows.

We demonstrate a three-node, all-fiber, quantum network that is supported by a low-noise, scalable, and automated clock distribution system. 
This is realized by distributing photon pairs in the telecommunication C-band ($\sim 1.5$~$\mu$m) simultaneously with strong optical ``clock" pulses in the telecommunication O-band ($\sim 1.3$~$\mu$m).
Specifically, light is distributed from a central node over two 11~km-length fibers to two end nodes.
The pulses used for clock distribution are created by bias switching a laser diode while the pulses generating the photon pairs, through spontaneous parametric down-conversion (SPDC), are carved from a continuous-wave laser by a Mach-Zehnder modulator.
Our setup uses in-house, high-bandwidth, and scalable electronics to generate 3.7~V (peak-to-peak) pulses with near-Gaussian distributions with durations as low as 47~ps and sub-ps timing jitter.
We quantify the effect of Raman scattering by measuring the coincidence-to-accidental ratio ($CAR$) of the distributed photon pairs.

We find that the clock distribution system reduces the $CAR$ from $77\pm 14$ to $42\pm 2$, which is still sufficient for high high-fidelity qubit distribution using our clock synchronization system.
Furthermore, we observe only 2~ps of timing jitter (over 1~minute of integration) between clocks at the central and end nodes, suggesting our method can be used for high-rate networks. 
Our results, collected using a free-running data acquisition and control system, demonstrates synchronized quantum networking in the C-band using an O-band clock, which is relevant for practical implementations of classical-quantum co-existing communications.

\section{Setup}
 
\begin{figure*}[h!]
  \begin{center}
  \includegraphics[width=\textwidth]{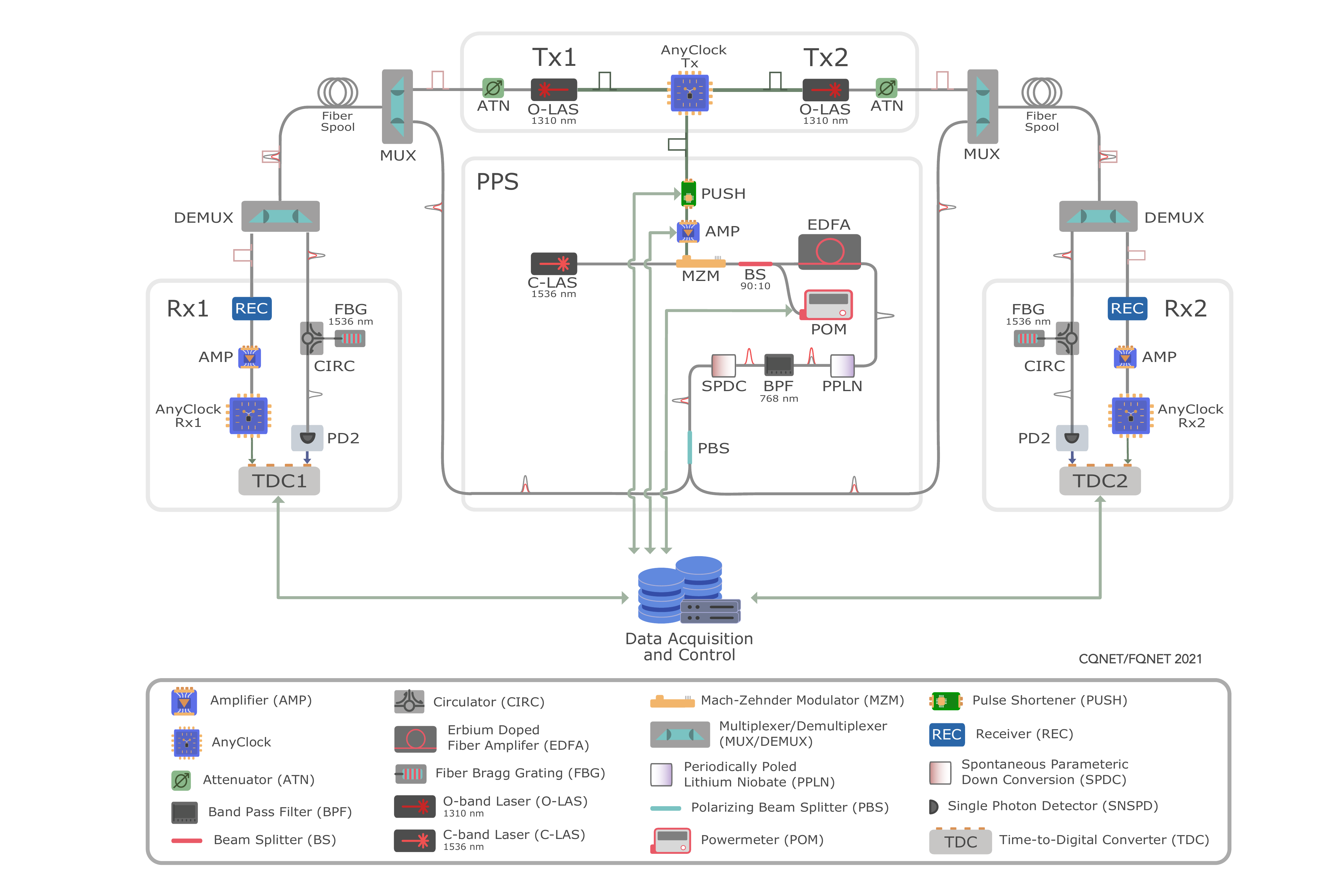}
  \caption{Schematic of fiber-based three-node quantum network and synchronization system. See main text for description. Clock pulses are indicated by top hats while grey and red Gaussian-shaped pulses indicate single photons and second-harmonic (768 nm) light, respectively.}
  \label{fig:setup}
  \end{center}
\end{figure*}
Our three-node quantum network and corresponding synchronization system is schematized in Fig.~\ref{fig:setup} and, other than the custom electronics and single-photon detectors, consists of fiber-based and off-the-shelf components.
The central node consists of a photon pair source (PPS) operating at the telecommunication C-band wavelength of 1536~nm and two transmitters (Tx1, Tx2) which generate clock pulses at the telecommunication O-band wavelength of 1310~nm.
By way of wavelength division multiplexer/demultiplexers (MUX/DEMUXs), the clock pulses and single photons are directed into fibers and distributed to end nodes via 11 km-length spools of single-mode fiber.
At the end nodes, the clock pulses and photon pairs are separated using DEMUXs and subsequently detected (Rx1, Rx2). 
As described in detail below, we ensure the photon pair generation and detection events are synchronized by (i) generating a photon pair synchronously with a clock pulse and (ii) recording the time between the detection of the clock pulse and the individual photon at each end node.

Clock distribution is seeded by a 200~MHz voltage oscillator (AnyClockTx) at the central node.
It generates 2.5~ns-duration pulses that are used to bias switch two O-band laser diodes (O-LASs), generating optical clock pulses of similar duration which are subsequently attenuated to an average power of 0.25~mW. 
This setup is indicated by Tx1 and Tx2 in Fig. \ref{fig:setup}.
Synchronous with the optical clock pulse generation, AnyClockTx creates a third pulse which is shortened to a duration as low as 47~ps (PUSH) and power-amplified by up to 28~dB (AMP) using customized electronics (see Sec. \ref{sec:electronics}), then directed to a 20~GHz-bandwidth fiber-coupled Mach-Zehnder modulator (MZM) within the PPS setup. The pulse amplitudes approximately correspond to the pi-voltage of the MZM.
The PPS setup contains a C-band laser (C-LAS) emitting continuous-wave light of 1536~nm wavelength that is modulated using the MZM to create 74 ps-duration optical pulses with an extinction ratio of 28~dB at a (clock-synchronized) repetition rate of 200~MHz. 
After passing a 90:10 beam splitter used for monitoring (POM) the stability of the MZM, these pulses are amplified by an erbium-doped fiber amplifier (EDFA), producing pulses with average power of 100~mW, and then directed to a fiber-packaged periodically poled lithium niobate (PPLN) waveguide which up-converts the light to 768~nm wavelength.
Next, residual 1536~nm light is removed using a band-pass filter (BPS) and the pulses are directed to a second PPLN waveguide configured to produce 1536~nm-wavelength photon pairs by Type-II SPDC.
A fiber-based polarizing beam splitter (PBS) separates each photon from the pair into different fibers, where they are each directed to the MUXs, and combined with the optical clock pulses in the fiber spools.

At the end nodes, after passing the DEMUXs, the individual photons are filtered by fiber Bragg gratings (FBGs), by way of circulators (CIRC), to a bandwidth of 2.5~GHz (see Sec. \ref{sec:discussion}), and detected using cryogenically cooled superconducting nanowire single photon detectors (SNSPDs) with 50~ps timing jitter (PD1 and PD2). 
The electrical pulses generated by the SNSPDs are directed to time-to-digital converters (TDC1, TDC2).
The optical clock pulses are received by 200~MHz-bandwidth amplified photodiodes (REC) which generate electrical pulses that are amplified (AMP) by 15~dB using scalable custom electronics (Sec. \ref{sec:electronics}).
These pulses adjust the phase of 200~MHz voltage oscillators (AnyClockRx1, AnyClockRx2) at the end nodes which produce pulses that are detected by the TDCs.
The TDCs then record the time difference between the electrical pulses generated by the SNSPDs and the oscillators to verify the synchronization.
This time difference is logged using a scalable data acquisition and monitoring system (indicated in Fig. \ref{fig:setup}) that enables uninterrupted quantum networking for an extended time duration ($>$days).

\section{Scalable and high-bandwidth custom electronics}
\label{sec:electronics}
Many pulsed quantum networking experiments use either mode-locked lasers \cite{valivarthi2016quantum} or laser diodes and MZMs driven by electrical pulses \cite{valivarthi2020teleportation} generated from arbitrary waveform generators \cite{tektronix} and off-the-shelf multi-purpose amplifiers~\cite{shf}.
These approaches are expensive, bulky, and not scalable to multi-node quantum networks.
We address this shortcoming by in-house developing custom pulse (duration) shorteners and amplifiers, referred to as Picoshort and Picoamp modules, respectively, that shape electrical pulses from the AnyClockTx oscillator.
See Fig. \ref{fig:setup}.
The resulting pulses are used to drive the MZM to its $\pi$-voltage, producing high-extinction pulses ($>$20~dB) for the PPS.
Short-duration ($<$100~ps) pulses allow the possibility of measuring high signal-to-noise ratios, the ability to create time-bin qubits in a single clock event (by splitting the pulse into two), and the realization of photon pairs with high spectral purity.
As shown in Fig. \ref{fig:setup}, we also use a Picoamp to increase the output voltage of the REC photodiodes for compatibility with the AnyClockRx oscillators.

\subsection{Pulse shortener (Picoshort)}
\begin{figure}
  \begin{center}
  \includegraphics[width=3.5in]{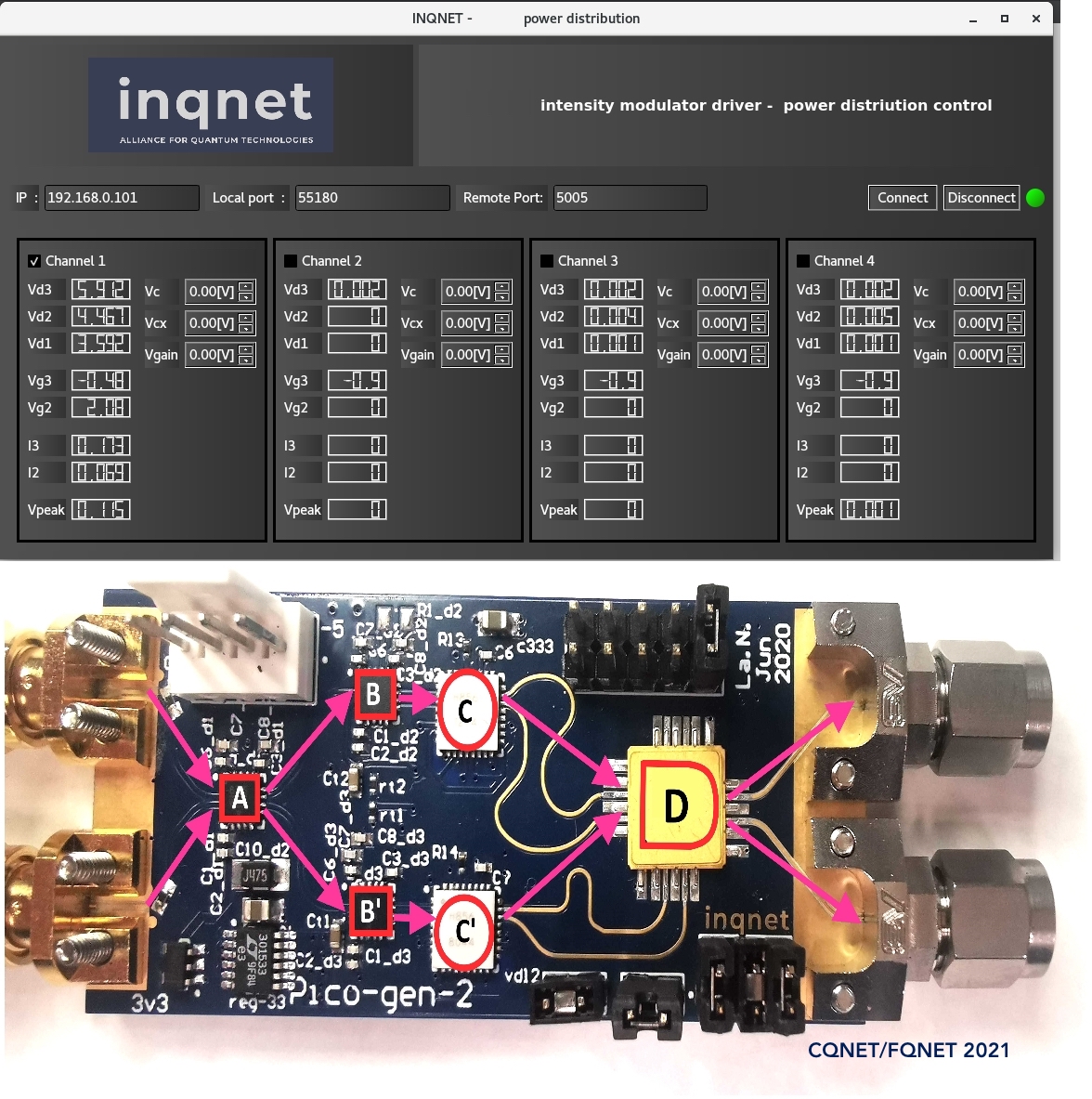}\\
  \caption{TOP: User interface to control the parameters of the mezzanine and Picoamp amplifier chip. 
  The interface displays the voltages and currents that are used to vary the pulse shape. BOTTOM: Picoshort pulse shortener board. The Picoshort receives a standard differential pulse and outputs a differential pulse as short as 25~ps. A: input discriminator, B,B': pair of discriminators with opposite logic output. C,C': variable delay lines. D: AND gate.}
  \label{fig:gui}
  \end{center}
\end{figure}

One viable approach to realize a pulse shortener is to use an AND gate combined with a variable delay line to produce electrical pulses with short ($<$100 ps) and tunable (e.g. from few to hundreds of ps) durations~\cite{garbati2017ultra}.
Accordingly we develop a circuit, the Picoshort, shown at the bottom of Fig.~\ref{fig:gui}, to shorten the 5~ns-duration pulses from the AnyClockTx oscillator. 
The input pulses are fed to a comparator (A) which produces differential digital signals with respect to a threshold. 
These signals are then directed to two comparators (B and B') which output differential digital signals with amplitudes that only differ in polarity. 
These pulses are then sent to two high bandwidth programmable variable delay lines (C and C') which have a resolution of 3~ps and a dynamic range of 100~ps, allowing the output pulse duration to be tuned.
The output pulses are then directed to an AND gate (D), which has a rise and fall time of 10~ps, thereby limiting the pulse duration to be no less than $\sim$25~ps.

\begin{figure}
  \begin{center}
  \includegraphics[width=3.5in]{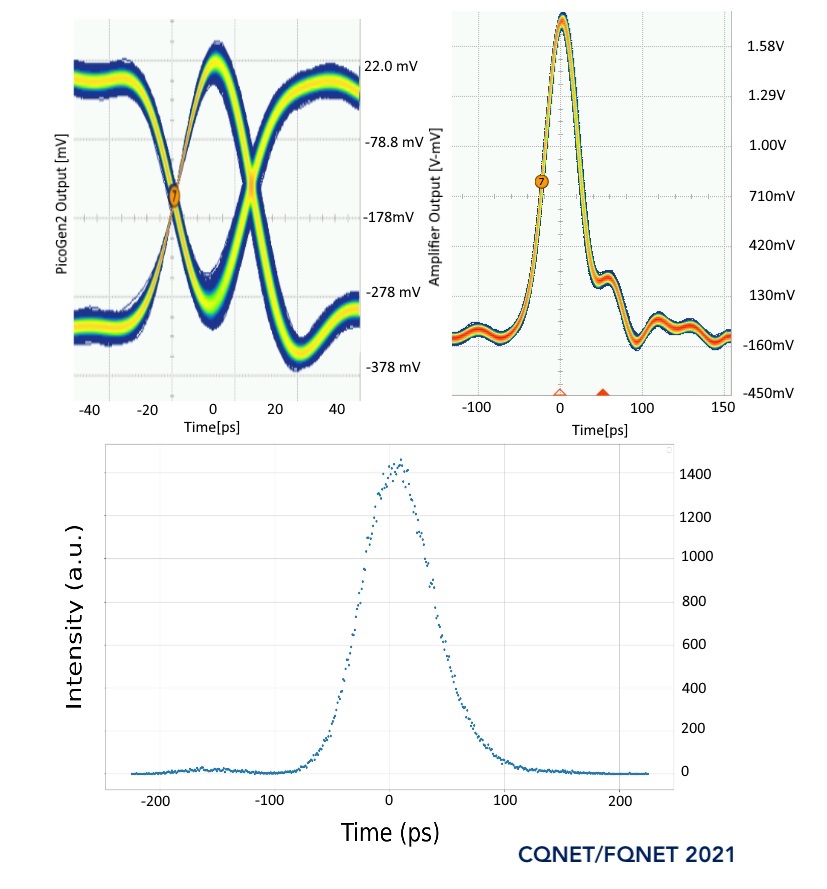}\\
  \caption{TOP LEFT: A 25~ps-duration differential output pulse from the Picoshort pulse shortener. 
  TOP RIGHT: A 47~ps-duration, 1.88~Vp-p, output pulse from the Picoamp amplifier.
  BOTTOM: Attenuated optical pulse measured after the MZM using an SNSPD. 
  Pulse duration is 74~ps and has an extinction ratio of 28~dB. Time is measured relative to the maximum amplitude of the pulse.}
  \label{fig:scope}
  \end{center}
\end{figure}

Using an input pulse from the AnyClockTx, the shortest duration output pulse from the Picoshort is measured using an oscilloscope. 
It is shown at the right side of Fig.~\ref{fig:scope} as a differential signal.
The full-width-at-half-maximum (FWHM) of the pulses is 24.8~ps and 24.5~ps for the positive- and negative-going differential signals, while their amplitudes are 270~mV and 299~mV, respectively.

\subsection{Pulse amplifier (Picoamp)}
\begin{figure}
  \begin{center}
  \includegraphics[width=3.5in]{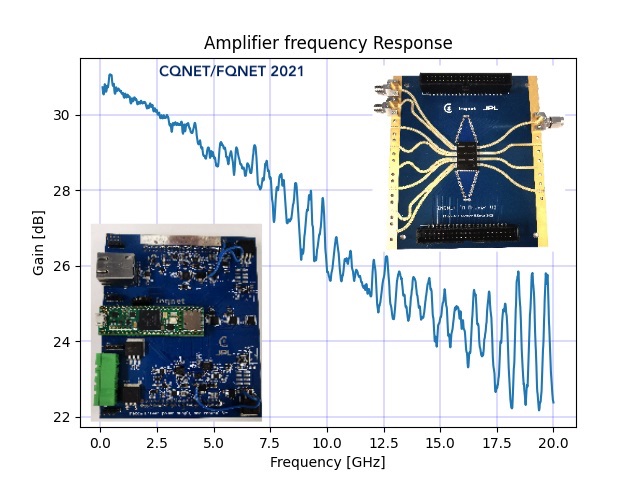}\\
  \caption{PLOT: Frequency response of the Picoamp differential-to-single-ended amplifier. RIGHT: Picoamp broadband differential-to-single-ended amplifier. LEFT: Mezzanine board for the Picoamp. It provides closed-loop control of the current and voltages and remote control of the tuning parameters via the interface shown in Fig. \ref{fig:gui}.}
  \label{fig:mcm_VNA_V3}
  \end{center}
\end{figure}

Amplification of the pulses from the Picoshort module is required to reach the $\sim$4~V $\pi$-voltage of the MZM to produce optical pulses of $<100$~ps duration. 
To this end, we create a high-bandwidth differential-to-single-ended radiofrequency amplifier board, the Picoamp, as shown in the right corner of Fig.~\ref{fig:mcm_VNA_V3}. 
A single board is able to drive four modulators simultaneously, allowing compact and inexpensive use in a future multi-node network or situations in which multiple modulators are required for generating the quantum signal \cite{liu2014mdiqkd, tang2015mdiqkd}.
As required for the Picoamp, we built an accompanying voltage biasing system on a mezzanine board shown in the left corner of Fig. ~\ref{fig:mcm_VNA_V3}.
The mezzanine board design facilitates seamless integration of our in-house-developed and user-friendly control software (screenshot shown at the top of Fig. \ref{fig:gui}) with the Picoamp for remote operation. 
The mezzanine and amplifier boards consist of several power supplies, current monitoring systems, analog-to-digital as well as digital-to-analog converters, and control the gain, zero-voltage crossing point, and the undershoot of the output pulses. 

To characterize the performance of the Picoamp, we first measure the frequency response (S$_{21}$) of one of its channels using a vector-network analyzer (Fig. \ref{fig:mcm_VNA_V3}).
We find a 3~dB-bandwidth of $\sim$10~GHz and a gain of more than 30~dB at low frequencies ($<$1~GHz), which is sufficient for amplifying pulses from the Picoshort.
The frequency-dependent modulation is likely due to resonances produced by imperfections in the board manufacturing, including track placement and connectors.

Next we amplify a 25~ps-duration pulse (that shown on the right side of Fig.~\ref{fig:scope}) from the Picoshort and, after optimizing for undershoot, voltage crossings, etc., we measure it using a oscilloscope, with result shown on the left side of Fig.~\ref{fig:scope}.
We find the amplitude and FWHM of the pulse to be 3.76~V and $\sim$47~ps, respectively, while the timing jitter of the pulse remains less than 1 ps.
The modulated tail of the pulse is likely due to small impedance mismatches within the Picoamp board.
Note that we measure similar timing jitters when using a Picoamp for increasing the voltage of pulses from the REC photodiode (Fig. \ref{fig:setup}).

Finally, we drive the MZM with the output pulse from the Picoamp.
For this step, we measure the output optical pulses after the MZM directly using a photodiode and oscilloscope, see the bottom of Fig. \ref{fig:scope}.
We find an optical pulse duration of 74 ps with an extinction ratio of 28 dB, with the slight pulse broadening owing to the ringing on the tail end of the pulse output from the Picoamp, the response of the MZM, and added jitter from the time-tagger.
Thus, we find that the Picoshort and Picoamp are a viable replacement to conventional bulky pulse generators, and it suitable for scalable quantum networks using PPSs.

\section{Results}

\begin{figure}
  \begin{center}
  \includegraphics[width=0.5\textwidth]{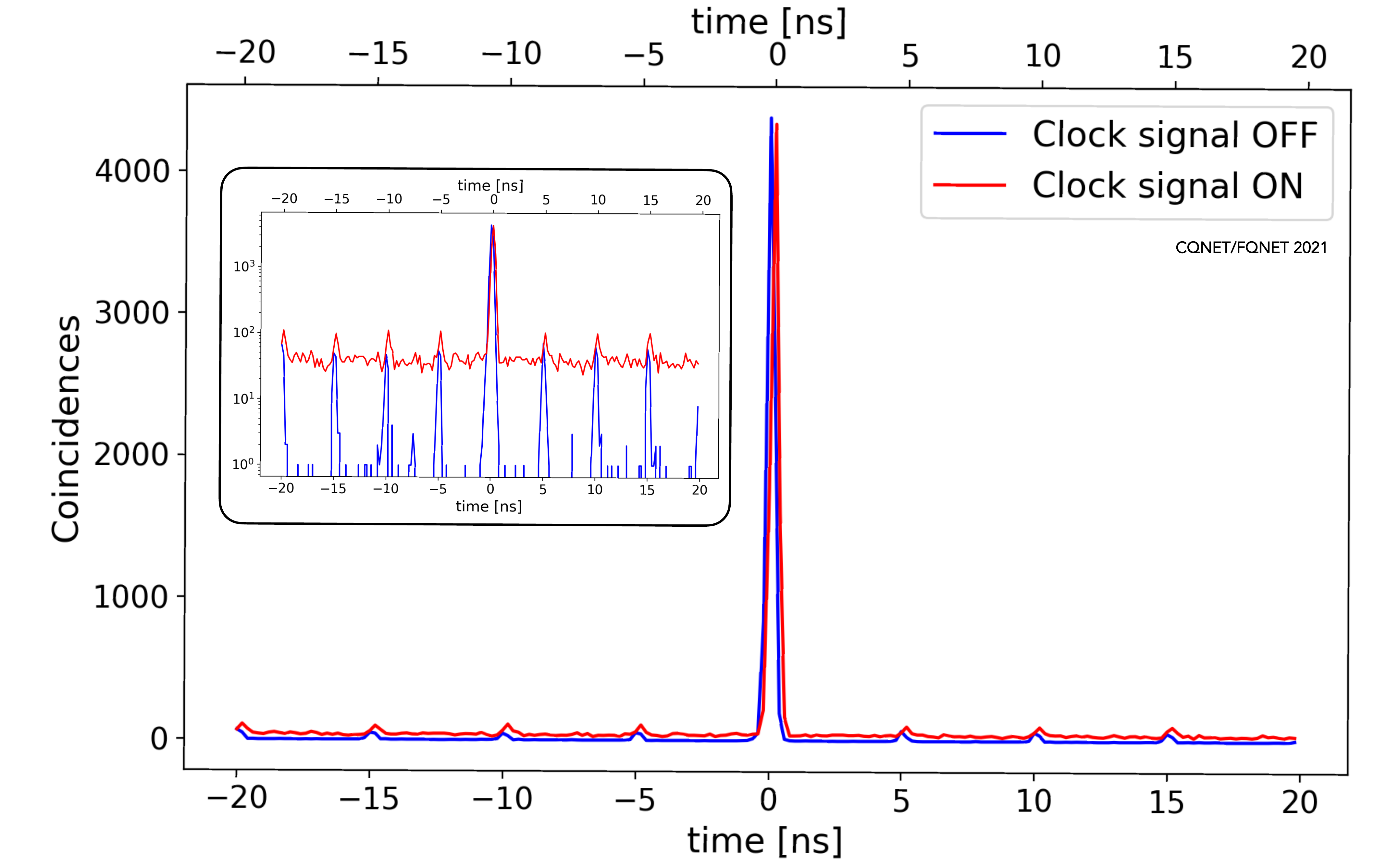}\\
  \caption{Coincidence histogram with the clock distribution enabled and disabled. The small time delay between the two histograms is due a small difference in trigger voltage threshold. Inset: Coincidence histogram with a log vertical scale reveals the Raman noise from the clock pulses.}
  \label{fig:CAR}
  \end{center}
\end{figure}

We characterize our quantum network setup by distributing photon pairs both with and without the clock distribution enabled, and measure the affect of the Raman noise from the clock pulses.
The role of noise is captured by the $CAR$ of the photon pairs, $CAR=C/A$, where, in the absence of noise, $C$ corresponds to the coincidence detection rate of photons originating from the same event, while $A$ corresponds to the coincidence detection rate of photons originating from different events.
Note that in this context, the $CAR$ is equivalent to the cross-correlation function $g^{(2)}(0)$ \cite{loudon2000}.
Dark counts and Raman scattering can reduce $CAR$ as a noise detection event may be recorded instead of a photon.
Our method is well-suited for quantum networks as channel loss ensures accidental coincidences.
Our PPS produces pairs with a probability ($\sim 1/CAR$) of $1\%$ per pulse~\cite{marcikic2002}, 
Channel loss, calculated by taking the ratio of the coincidence rates to the single photon detection rates~\cite{marcikic2002}, from PPLN waveguide to the SNSPDs are 24~dB and 26~dB.
We measure the arrival time difference of SNSPD detection events at Rx1 and Rx2 over 5~minutes, both with and without the clock pulses, compiling all events into histograms, see Fig. \ref{fig:CAR}.
We sum the detection events over a 200~ps interval around the peak at zero time delay to calculate $C$, while the average number of detection events in a 200 ps interval around each of the accidental peaks is used to determine $A$.
The measurements yield a CAR of $77\pm 12$ without and a CAR of $42\pm 2$ with the clock distribution enabled, respectively, which are both well above the classical limit of $CAR=2$.
To place our results into context for qubit distribution, if these photon pairs were to be time-bin entangled, e.g. using the approach demonstrated in Ref. \cite{valivarthi2020teleportation}, and if no other imperfections play a role, the measured reduction of $CAR$ from the Raman noise suggests a reduction of fidelity $CAR/(CAR+1)$ \cite{takesue2005a} from 99$\%$ to 98$\%$ or the same reduction in the entanglement visibility $(CAR-1)/(CAR+1)$ \cite{takesue2005b}.
This visibility is well-above the $1/3$ required for non-separability of a Werner state \cite{werner1989} and the non-locality bound of $1/\sqrt{2}$ \cite{clauser1969}. 
Thus, the noise introduced by our clock distribution system plays a minimal role in a quantum network.

\begin{figure}
  \begin{center}
  \includegraphics[width=3.5in]{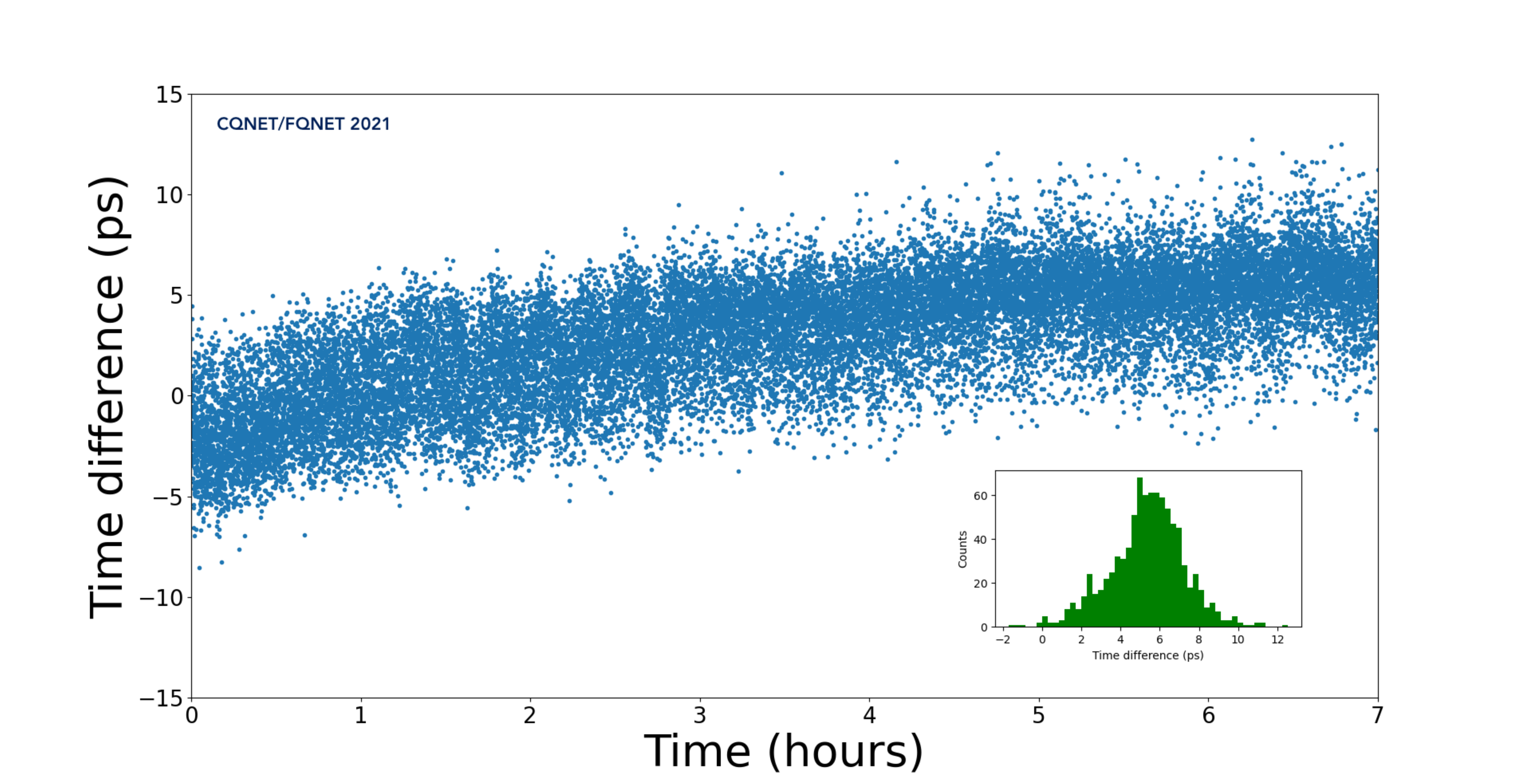}\\
  \caption{Variation of the time difference between the arrival of clock pulses at Rx1 and Rx2 over 7 h. The maximum time difference is 5 ps due to fiber length variations. Inset: histogram of the time difference over a 900 s time scale indicates a timing jitter of 2 ps.}
  \label{fig:jitter}
  \end{center}
\end{figure}

Importantly, we also determine the timing jitter of our clock distribution method, which sets an upper-bound on the rate of the quantum network.
To compare the arrival time of the clock pulses at Rx1 and Rx2 (after the AnyClockRx1 and AnyClockRx2 oscillators), we use an oscilloscope to measure a timing jitter of 2~ps over a timescale of 60~s, and a time difference that slowly drifts by 5~ps over 7~h owing to fiber length variations, see Fig. \ref{fig:jitter}.
Note that we use an oscilloscope because the current configuration of the TDC adds up to 7~ps timing jitter, but with a standard upgrade this can be as low as 3~ps \cite{qutag}.
Since the clock pulses are attenuated to ensure a minimal reduction in $CAR$, our measurement is limited by the noise floor of detectors only.
Nevertheless, the timing jitter of our clock distribution currently sets an upper-bound on our distribution rate of $\sim300$~MHz, which is sufficient for long-distance networks.  

\section{Discussion and outlook}
\label{sec:discussion}
Despite not constituting the optimal choice of wavelength, our telecommunication O-band synchronization system introduces little noise into our telecommunication C-band quantum network. The low noise is partially due to the strong spectral filtering of the photons at the FBGs -- a required step to ensure the photons are purified, i.e. spectral correlations are removed.
This renders the photons suitable for interference, as required for implementations of advanced network protocols, e.g. based on quantum teleportation.
Note that the 768~nm light remaining after the SPDC step is far off-resonant from the 1536~nm photons, and is partially filtered by the long fiber spools, MUX/DEMUX filters, fiber Bragg gratings, and SNSPD devices, thus it does not contribute any measurable noise to the network.
Further reduction of noise in our system can be afforded by detecting the clock pulses with more sensitive REC detectors, thus allowing a reduction of the clock pulse intensity.
To this end, SNSPDs operating in the O-band could be used, and would constitute minimal system overhead given that C-band SNSPDs are already deployed.
This would also result in improvements to the system clock rate as SNSPDs feature timing jitters as low as a few ps \cite{korzh2020lowjitter}, which constitutes an upper-bound of a few hundred GHz to the clock rate (note that the impact of the dead time of the SNSPDs is negligible due to channel loss).
Although this rate cannot be reached by our current implementation, we expect that clock rates of a few GHz can be achieved simply by using a GHz-bandwidth REC photodiode and exchanging our AnyClocks with stabilized GHz-frequency oscillators and accompanying phase control circuits.
To ensure there is no reduction in CAR at these rates, it is likely fine-tuning of the impedance matching of electronic circuits and between components is necessary, along with using higher-bandwidth FBG filters and MZMs.
In addition, the CAR could be improved if the clock pulses are transmitted in the telecommunication L-band ($\sim$1.6~$\mu$m), or if the PPS operated at a wavelength that is blue shifted from the clock, e.g. in the O-band while operating the clock in the C-band.
Nonetheless, the C-band features the lowest loss in standard single-mode fiber used for long-haul networks (0.18~dB/km), while the O-band is a historically and commonly used channel, and constitutes an operating wavelength that matches those of readily available off-the-shelf telecommunication components (e.g. lasers, modulators, MUX/DEMUXs, detectors, etc.).
Overall, our three-node quantum network and accompanying synchronization system sheds light on the role of noise in quantum networking and constitutes a step towards practical, as well as  high-rate, classical-quantum co-existing networks.

\section*{Acknowledgment}
We acknowledge partial funding from the Department of Energy BES HEADS-QON Grant No. DE-SC0020376 (targeting tranduction work), QuantiSED SC0019219, and the AQT Intelligent Quantum Networks and Technologies (INQNET) research program.  Partial support for this work was provided by the Caltech/JPL PDRDF program.  S.I.D. and A.M. acknowledge partial support from the Brinson Foundation.  Part of this research was performed at the Jet Propulsion Laboratory, California Institute of Technology, under contract with NASA.  C.P. further acknowledges partial support from the Fermilab's Lederman Fellowship and LDRD. D.O. acknowledges support of the Natural Sciences and Engineering Research Council of Canada through the Discovery Grant program and the CREATE QUANTA training program, and the Alberta Ministry for Jobs, Economy and Innovation through the Major Innovation Fund Quantum Technologies Project (QMP). We are grateful  to  Jason Trevor (Caltech Lauritsen Laboratory for High Energy Physics),  Vikas Anant (PhotonSpot) as well as Artur Apresyan and the HL-LHC USCMS-MTD Fermilab group.



\end{document}